\documentclass[12pt,oneside]{article}
\usepackage{epsfig, cite}
\usepackage{color}
\usepackage{ifthen}
\usepackage{graphicx}

\def\p{\partial}

\def\half{{1\over 2}}
\def\({\left(}
\def\){\right)}
\def\[{\left[}
\def\]{\right]}

\def\p{\partial}

\def\half{{1\over 2}}
\def\({\left(}
\def\){\right)}
\def\[{\left[}
\def\]{\right]}

\def\e{\begin{equation}}
\def\q{\end{equation}}
\def\m{\begin{eqnarray}}
\def\n{\end{eqnarray}}

\begin{document}
\thispagestyle{empty} \setcounter{page}{0}

\vspace{2cm}

\begin{center}
{\Large Weak gravity conjecture constraints on inflation}

\vspace{1.4cm}

Qing-Guo Huang

\vspace{.3cm}

{\em School of physics, Korea Institute for
Advanced Study,} \\
{\em 207-43, Cheongryangri-Dong, Dongdaemun-Gu, } \\
{\em Seoul 130-722, Korea}\\
\end{center}

\vspace{-.1cm}

\centerline{{\tt huangqg@kias.re.kr}} \vspace{1cm}
\centerline{ABSTRACT}
\begin{quote}
\vspace{.5cm}

We consider the gravitational correction to the coupling of the
scalar fields. Weak gravity conjecture says that the gravitational
correction to the running of scalar coupling should be less than the
contribution from scalar fields. For instance, a new scale
$\Lambda=\lambda_4^{1/2}M_p$ sets a UV cutoff on the validity of the
effective $\lambda_4 \phi^4$ theory. Furthermore, this conjecture
implies a possible constraint on the inflation model, e.g. the
chaotic inflation model might be in the swampland.

\end{quote}
\baselineskip18pt


\vspace{5mm}

\newpage

\setcounter{equation}{0}

A full quantum theory of gravity has not been formulated. It is a
good idea to work out some insights coming from the marriage of
gravity and quantum mechanism. Usually we quantize matter field and
gravity separately. However, many authors in
\cite{Vafa:2005ui,Arkani-Hamed:2006dz} suggested that gravity and
the other gauge forces cannot be treated independently. A nontrivial
constraint on the really self-consistent gauge theories with gravity
appears.

In \cite{Arkani-Hamed:2006dz}, a new intrinsic UV cutoff for a
four-dimensional U(1) gauge theory $\Lambda\sim gM_p$ is proposed,
where $g$ is the U(1) gauge coupling and $M_p$ is four-dimensional
Planck scale. This conjecture is called ``weak gravity conjecture''
and it provides a possible criterion on the string landscape.
However, there are a huge number of meta-stable vacua in string
landscape with positive vacuum energy. The authors in
\cite{Huang:2006hc} proposed a weak gravity conjecture for the case
with a cosmological constant. In \cite{Huang:2006pn}, we showed that
it is possible to check this conjecture at the LHC if there are
several large extra dimensions and the fundamental Planck scale is
roughly 1 TeV. This conjecture has also been generalized to higher
dimensions \cite{Banks:2006mm,Huang:2007mf}. In particular, the
author in \cite{Huang:2007mf} gave a new insight on weak gravity
conjecture for gauge theories at the one-loop level and suggested a
new argument which is independent on monopole and black hole. In
\cite{Huang:2006tz}, the authors proposed a weak gravity conjecture
for noncommutative field theories. Other related works are discussed
in
\cite{Kachru:2006em,Kats:2006xp,Li:2006vc,Adams:2006sv,Ooguri:2006in,Li:2006jj,Medved:2006ht,Gasperini:2006as}.

Guth in \cite{Guth:1980zm} suggested a possiblity to solve the
horizon, flatness and primordial monopole problem due to a
quasi-exponential expansion (inflation) of the universe before hot
big bang. However his inflation model called old inflation model
leads to an extremely large inhomogeneity and anisotropy of the
universe after the phase transition \cite{Guth:1982pn}. In
\cite{Albrecht:1982wi,Linde:1982zj}, a new version of inflation to
solve previous problems in old inflation model was proposed where
inflation is governed by the potential of the scalar fields.

It is well-known that the gauge interaction is determined by
symmetry. Unfortunately, there is not a physical principle to
constrain the interaction of the scalar fields. In this paper, we
generalize the conjecture in \cite{Huang:2007mf} to scalar field
theories and investigate the constraints on the inflation model.

The interactions between graviton and the scalar fields in $d$
dimensions are described by the action \e S={1\over 2\kappa_d^2}\int
d^dx \sqrt{-g} R+\int d^dx \sqrt{-g}\( \half g^{mn}\p_m \phi \p_n
\phi+\half m^2 \phi^2 +V(\phi)\), \label{gsc}\q where
$\kappa_d^2\sim G_d$ is the $d$-dimensional Newton coupling constant
and $V(\phi)$ is the potential of the scalar field. In order to work
out the coupling between graviton and scalar field, we consider the
quantum fluctuation of the gravitational degrees of freedom around
the Minkowski metric as \e g_{mn}=\eta_{mn}+\kappa_d h_{mn}. \q The
action (\ref{gsc}) becomes \e S\sim \int d^dx (\p h)^2 + \int d^dx
\half \kappa_d \p_m \phi \p_n \phi h^{mn}+\cdot \cdot \cdot.\q The
interaction term between graviton and the scalar field is
proportional to positive powers of $\kappa_d$.

In this note we focus on the scalar field theories with polynomial
potential. For the potential of the scalar field  \e
V(\phi)=\lambda_n\phi^n,\label{pt}\q the scalar field $\phi$ has
dimensionality $[\phi]=[mass]^{d-2\over 2}$ and the coupling
$\lambda_n$ takes $[\lambda_n]=[mass]^{d-{n(d-2)\over 2}}$. In
general the non-renormalizable interactions are just those whose
coupling constants have the dimensionality of negative powers of
mass. In $d$ dimensions, $\lambda_n\phi^n$ term for the scalar field
theories are non-renormalizable if $n>{2d\over d-2}$. For $d=4$, the
interaction terms with $n\leq 4$ are renormalizable. In this paper,
we only consider the scalar field theories in four dimensions and
our results can be easily generalized to higher dimensions. We do
not consider three or lower dimensions as gravity does not contain
propagating degrees of freedom in these dimensions, even though some
of our conjectures may be applicable to these cases.

In \cite{Robinson:2005fj}, Robinson and Wilczek calculated the
gravitational correction to running of gauge couplings in four
dimensions. In \cite{Huang:2007mf}, the author gave several examples
to support a new viewpoint of weak gravity conjecture which says
that the gravitational correction to the $\beta$ function should be
less than the contribution from the gauge fields. In this short
note, we generalize the observations in \cite{Huang:2007mf} to
scalar field theories.

First, we consider the scalar field theory with potential \e
V(\phi)=\lambda_4 \phi^4.\q The typical Feynman diagrams
contributing to the running of the scalar coupling $\lambda_4$ are
showed in Fig. 1.
\begin{figure}[h]
\begin{center}
\leavevmode \epsfxsize=0.6\columnwidth \epsfbox{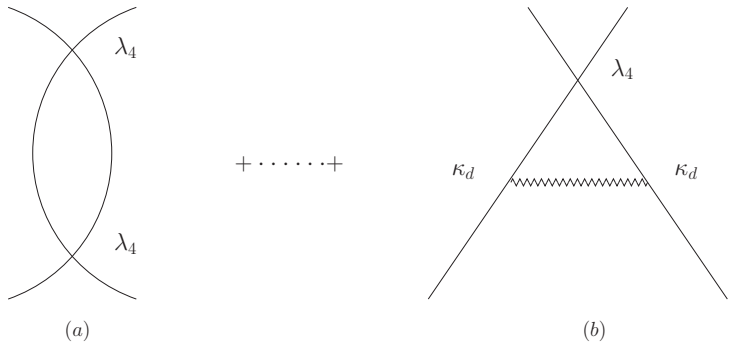}
\end{center}
\caption{The typical Feynman diagrams for a scalar field
contributing (a) and a gravitational contributing (b) to the $\beta$
function at one-loop level.}
\end{figure}
The dimensional analysis reads the $\beta$ function for $\lambda_4$
\e \beta_4={d\lambda_4\over d\ln \Lambda}\sim
c_4\lambda_4^2-c_l\lambda_4\({\Lambda\over M_P}\)^2,\label{btf} \q
here $c_4$ and $c_l$ are the numerical coefficients, and $\Lambda$
is an energy scale. When Planck scale is much larger than the scale
for the field theory, the gravitational correction can be ignored.
Generically the only useful and meaningful notion of the RG running
is related to the logarithm corrections to the coupling constant.
The power divergence in eq. (\ref{btf}) signals that there is an
intrinsic uncertainty in the field theory predictions due to the
presence of the higher dimensional operators induced by the
gravitational corrections. In principle, we cannot precisely
calculate the gravitational corrections before we know the full
quantum theory of gravity. We expect that any would-be fundamental
theory of quantum gravity should reproduce the same result in the
limit of the physical scenario considered here.

The weak gravity conjecture for U(1) gauge theory is motivated by
the absence of the global symmetry for the quantum gravity
\cite{Arkani-Hamed:2006dz}. Actually this conjecture is consistent
with the requirement that the gravity should be the weakest force.
This can be also read out from the $\beta$ function
\cite{Huang:2007mf}. Therefore we generalize this idea to the scalar
field theory. Requiring that the contribution from scalar fields is
greater than gravitational correction yields \e \Lambda^2 \leq
\lambda_4M_P^2. \label{wcf} \q Beyond the scale $\lambda_4^\half
M_p$, the effective $\lambda_4\phi^4$ theory breaks down and a full
quantum theory including gravity is needed. If the mass square is
negative, this scalar field plays the role as Higgs field. The
vacuum expectation value (vev) of $\phi$ is $\langle\phi\rangle\sim
\sqrt{ m^2/\lambda_4}$ which is nothing but the electro-weak scale
in standard model. The mass of the physical scalar field around
$\langle\phi\rangle$ is still $m$. If the UV cutoff of the scalar
field theory $\Lambda$ is smaller than the mass of the scalar field,
any quanta of this theory cannot be excited. Naturally $m\leq
\Lambda$ is required. Substituting this inequality into eq.
(\ref{wcf}), we find \e \langle\phi\rangle\sim \sqrt{m^2\over
\lambda_4}\leq M_p, \label{wcfm}\q which says that the electro-weak
scale should be lower than Planck scale. Eq. (\ref{wcfm}) is also
conjectured in \cite{Li:2006jj} where the authors proposed a piece
of evidence in two dimensions. It is a reasonable, but trivial
observation with the viewpoint of field theory. However there will
be a significant implication for inflation model.

Before we consider the weak gravity conjecture constraints on
inflation, we investigate another example with potential \e
V(\phi)=\lambda_3 \phi^3. \q In principle, we cannot well define
this field theory, since its potential has not lower bound. Here we
assume that there is still such an effective field theory and the
possible Feynman diagrams contributing to the $\beta$ function are
showed in Fig. 2.
\begin{figure}[h]
\begin{center}
\leavevmode \epsfxsize=0.6\columnwidth \epsfbox{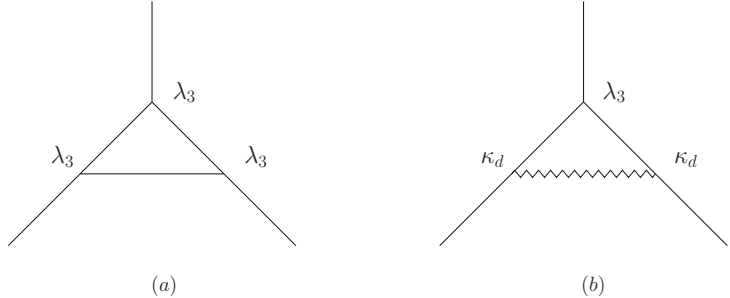}
\end{center}
\caption{The typical Feynman diagrams for a scalar field
contributing (a) and a gravitational contributing (b) to the $\beta$
function at one-loop level.}
\end{figure}
Dimensional analysis reads that the $\beta$ function takes the form
\e \beta_3={d\lambda_3\over d\ln \Lambda}\sim
c_3\lambda_3^3\Lambda^{-2}-c_l\lambda_3\({\Lambda\over M_P}\)^2, \q
where $c_3$ and $c_l$ are the numerical coefficients. Requiring the
contribution from scalar field is greater than gravitational
correction yields \e \Lambda^2 \leq \lambda_3M_P. \label{wct} \q
Define a dimensionless coupling ${\tilde \lambda}_3=\lambda_3/M_p$.
Eq. (\ref{wct}) is expressed as $\Lambda\leq {{\tilde
\lambda}_3}^{1/2}M_p$.

Following we switch to use the weak gravity conjecture for the
scalar field theories to constrain the inflation models. In the
slow-roll inflation model, the potential of scalar field provides an
effective positive cosmological constant $\rho_V\sim V(\phi)$. The
Hubble parameter is governed by Friedmann equation \e H^2\sim
{\rho_V\over M_P^2}.\q Hubble constant acts as the IR cutoff for the
field theories. Naturally the IR cutoff of a field theory is lower
than the UV cutoff \cite{Huang:2006hc}, i.e. \e H \leq \Lambda.
\label{sf}\q This is nothing but that the size of the universe
$H^{-1}$ is larger than the shortest physical length of the field
theories $1/\Lambda$. On the other hand, the quantum fluctuations of
the scalar field on this quasi de Sitter background takes the form
\e \langle \phi^2 \rangle \sim H^2. \q Eq. (\ref{sf}) can be
interpreted as the quantum fluctuation of the scalar field is
smaller than UV cutoff of the field theory.

The evolution of slow-roll inflation is dominated by the potential
of scalar field. For instance, we consider the case with $n=4$ in
(\ref{pt}). If $\lambda_4\phi^4$ term dominates the potential, the
Hubble constant is related to the scalar field by \e H^2\sim
\lambda_4\phi^4/M_P^2. \q Taking into account the weak gravity
conjecture (\ref{wcf}) and (\ref{sf}) yields \e
{\lambda_4\phi^4\over M_P^2}\leq \Lambda^2\leq \lambda_4M_P^2, \quad
\hbox{or}, \quad \phi\leq M_P. \label{wcfl}\q The value of scalar
field should be smaller than the Planck scale, which is just what we
expect. On the other hand, for the case with the evolution of the
universe governed by the mass term, i.e. $m^2\phi^2>\lambda_4\phi^4$
, or, \e \phi^2\leq {m^2\over \lambda_4},\q the weak gravity
conjecture (\ref{wcfm}) implies $\phi\leq M_p$ as well. Similarly,
we also get $\phi\leq M_P$ for $n=3$. We need to remind the readers
that in chaotic inflation model \cite{Linde:1983gd} the value of the
scalar field is roughly $\sqrt{N}M_p$ which is greater than the
four-dimensional Planck scale, where $N$ is the number of e-folds
before the end of inflation. Weak gravity conjecture implies that
chaotic inflation might be in the swampland where the theory is just
semi-classically consistent, but not actually inconsistent on the
quantum level.

For the scalar field theory with the potential \e
V(\phi)=V_0+\lambda_4\phi^4, \q where $V_0$ is constant and acts as
a zero point energy in field theory. First we ignore the
contribution to the $\beta$ function from $V_0$. Now the prediction
of the weak gravity conjecture (\ref{wcf}) is still valid. If
$\lambda_4\phi^4$ term dominates the evolution of inflation, the
inflation is just $\lambda_4\phi^4$ inflation and the results have
been showed previously. Now we consider the case when the inflation
is dominated by $V_0$, i.e. \e V_0\geq \lambda_4\phi^4. \label{vf}\q
The Hubble constant takes the form \e H^2\sim V_0/M_P^2. \q Weak
gravity conjecture implies that $H^2\leq\Lambda^2\leq
\lambda_4M_P^2$, or \e V_0\leq \lambda_4M_P^4. \label{wcv}\q
Combining with (\ref{vf}), we find $\phi\leq M_p$ again. In this
inflation model the slow-roll conditions can be satisfied even when
the value of the inflaton is smaller than $M_p$. But the weak
gravity conjecture still brings a constraints on the value of scalar
field at the end of inflation $\phi_e$, since the total number of
e-fold during inflation is greater than 60 at least. The value of
the inflaton $\phi_N$ at the number of e-folds $N$ before the end of
inflation is related to the value of the inflaton at the end of
inflation $\phi_e$ by \e N\sim {V_0\over M_p^2\lambda_4} \({1\over
\phi_e^2}-{1\over \phi_N^2} \). \label{np} \q This potential of
inflaton is a typical potential for hybrid inflation model where the
inflation cannot be ended by violating the slow-roll conditions
\cite{Linde:1993cn}. Combing (\ref{np}) with (\ref{wcv}) implies \e
{N\over M_p^2}+{1\over \phi_N^2}\leq {1\over \phi_e^2}. \q Taking
$\phi_N\leq M_p$ into account, we find \e \phi_e\leq
M_p/\sqrt{N+1}.\q In order to solve the flatness problem in the hot
big bang model, the number of e-folds should not be less than 60 and
then $\phi_e\leq 0.13 M_p$; or, the total number of e-folds has an
upper bound $N_{total}\leq M_p^2/\phi_e^2-1$.

We also want to consider the case with potential
$V=\lambda_6\phi^6$. If we only consider $\lambda_6\phi^6$ term,
this theory is non-renormalizable and the $\phi^4$ counter term
should be naturally included. Usually we don't tune the coefficient
of $\phi^4$ to be zero in an effective field theory. Therefore we
consider the scalar field theory with potential \e
V(\phi)=\lambda_4\phi^4+\lambda_6\phi^6. \label{pfs}\q The $\beta$
function for $\lambda_4$ at the one-loop level takes the form \e
\beta_4={d\lambda_4\over d\ln \Lambda}\sim
c_4\lambda_4^2+c_6\lambda_6\Lambda^2-c_l\lambda_4\({\Lambda\over
M_p}\)^2, \label{bfs}\q where $c_6$ is the numerical coefficient and
the middle term on the right hand side of (\ref{bfs}) comes from
Fig. 3.
\begin{figure}[h]
\begin{center}
\leavevmode \epsfxsize=0.2\columnwidth \epsfbox{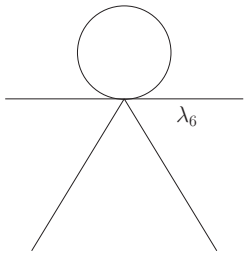}
\end{center}
\caption{The typical Feynman diagram contributing to $\beta_4$ from
$\lambda_6\phi^6$ at one-loop level.}
\end{figure}
Naively weak gravity conjecture implies that both the first and the
second term on the right hand side of (\ref{bfs}) should be greater
than the gravitational correction respectively, i.e. \e
{\Lambda^2\over M_p^2}\leq \lambda_4\leq \lambda_6M_p^2.
\label{wcfs}\q If $\lambda_4\phi^4$ term dominates inflation, the
results are just the same as the case with only $\lambda_4\phi^4$.
If inflation is driven by $\lambda_6\phi^6$, the Hubble constant is
$H^2\sim \lambda_6\phi^6/M_p^2$. We rewrite eq. (\ref{wcfs}) as \e
\Lambda^2\leq \lambda_6M_p^4. \label{swc}\q This result can be
obtained as well from the $\beta$ function for $\lambda_6$.
Dimensional analysis says that $\beta$ function for $\lambda_6$ at
the two loop level takes the form $\beta^{2-loop}_6\sim
v_6\lambda_6^2\Lambda^2-v_l\lambda_6{\Lambda^4\over M_p^4}$.
Requiring the two-loop gravitational correction is less than the
two-loop scalar contribution yields eq. (\ref{swc}). Now $H\leq
\Lambda$ also leads to $\phi\leq M_p$. This chaotic inflation model
might be in the swampland as well.

The previous arguments can be generalized to the cases with multi
fields. To be simple, we consider the assisted inflation model
\cite{Liddle:1998jc} with \e V(\phi_i)=\lambda_4\phi_i^4, \quad
i=1,...,n.\q A possible assisted chaotic inflation model in string
theory was proposed in \cite{Dimopoulos:2005ac}. In assisted
inflation, there is a unique late-time attractor with all the scalar
fields equal, i.e. $\phi_1=\phi_2=...=\phi_n$, where $n$ is the
number of scalar fields. With this ansatz, the Hubble constant is
given by \e H^2\sim {n\over M_p^2}\(\half
m^2\phi^2+\lambda_4\phi^4\),\q where $\phi=\phi_i$. The equation of
motion for $\phi$ is \e 3H\dot \phi\simeq -{dV(\phi)\over d\phi}.\q
The number of e-folds $N$ before the end of inflation is related the
value of $\phi_N$ by \e N\sim \int Hdt\sim -{n\over
M_p^2}\int_{\phi_N}^{\phi_e}{V\over V'}d\phi. \label{mnt}\q Since
there is no coupling among these scalar fields from the viewpoint of
the field theory, the weak gravity conjecture (\ref{wcf}) is still
valid for each scalar field $\phi$. If $\lambda_4\phi^4$ dominates
the evolution of inflation, (\ref{wcfl}) is replaced by \e
{n\lambda_4\phi_N^4\over M_p^2}\leq \Lambda^2\leq \lambda_4M_p^2.
\label{mwcf}\q Using eq. (\ref{mnt}), the value of $\phi$ at the
number of e-folds $N$ before the end of inflation takes the form
$\phi_N\sim \sqrt{N/n}M_p$. Now eq. (\ref{mwcf}) reads \e N\leq
\sqrt{n}.\q Requiring that the total number of e-folds is not less
than 60 implies that the number of inflaton is larger than $3600$.
On the other hand, the total number of e-folds is bounded by the
number of the inflatons. Similar results are also obtained for the
case with inflation dominated by $m^2\phi^2$ term.

To summarize, we propose a weak gravity conjectures for scalar field
theories case by case, because there is not a principle to determine
the shape of the potential for scalar fields. We only investigate
the scalar field theories with polynomial potential and we find a
free scalar field theory can not be self-consistent involving
gravity. We expect that our results do not rely on perturbation
theory, but we usually need to use perturbation theory to calculate
the $\beta$ function.

We also use the weak gravity conjecture for the scalar field
theories to constrain inflation models. Usually we take the Planck
scale as a natural cutoff for the field theories and the field
theories do not break down as long as the energy density is smaller
than the Planck density. But weak gravity conjecture suggests that
gravity cannot be ignored at a new scale lower than Planck scale. As
a result, the value of the inflaton should be smaller than Planck
scale for chaotic inflation and then we say that chaotic inflation
model might not be realized in an effective field theory. Here we
need to stress that we cannot give a general argument to support
that the value of the inflaton cannot be greater than Planck scale
for arbitrary potential. In this paper, we only propose one
condition $H\leq \Lambda$ to constrain inflation model. More
insights are called for if one wants to get more stringent
constraints on inflation model. In fact, our results are also valid
for quintessence \cite{Zlatev:1998tr}, a candidate for dark energy.

Actually our intriguing proposals are still less certain. But we
hope that our observations can be taken as the first step on the
line to the correct answer.

\vspace{.5cm}

\noindent {\bf Acknowledgments}

We would like to thank P.M. Ho, J. Lee, M. Li and F.L. Lin for
useful discussions. We also thank department of physics in National
Taiwan University for the hospitality.


\newpage

\begin{thebibliography}{99}
\baselineskip=16pt

\bibitem{Vafa:2005ui}
  C.~Vafa,
  ``The string landscape and the swampland,''
  arXiv:hep-th/0509212.

\bibitem{Arkani-Hamed:2006dz}
  N.~Arkani-Hamed, L.~Motl, A.~Nicolis and C.~Vafa,
  ``The string landscape, black holes and gravity as the weakest force,''
  arXiv:hep-th/0601001.

\bibitem{Huang:2006hc}
  Q.~G.~Huang, M.~Li and W.~Song,
  ``Weak gravity conjecture in the asymptotical dS and AdS background,''
  JHEP {\bf 0610}, 059 (2006)
  [arXiv:hep-th/0603127].

\bibitem{Huang:2006pn}
  Q.~G.~Huang,
  ``Weak Gravity Conjecture with Large Extra Dimensions,''
  arXiv:hep-th/0610106.

\bibitem{Banks:2006mm}
  T.~Banks, M.~Johnson and A.~Shomer,
  ``A note on gauge theories coupled to gravity,''
  JHEP {\bf 0609}, 049 (2006)
  [arXiv:hep-th/0606277].

\bibitem{Huang:2007mf}
  Q.~G.~Huang,
  ``Gravitational Correction and Weak Gravity Conjecture,''
  JHEP {\bf 0703}, 053 (2006)
  [arXiv:hep-th/0703039].

\bibitem{Huang:2006tz}
  Q.~G.~Huang and J.~H.~She,
  ``Weak gravity conjecture for noncommutative field theory,''
  JHEP {\bf 0612}, 014 (2006)
  [arXiv:hep-th/0611211].

\bibitem{Kachru:2006em}
  S.~Kachru, J.~McGreevy and P.~Svrcek,
  ``Bounds on masses of bulk fields in string compactifications,''
  JHEP {\bf 0604}, 023 (2006)
  [arXiv:hep-th/0601111].

\bibitem{Li:2006vc}
  M.~Li, W.~Song and T.~Wang,
  ``Some low dimensional evidence for the weak gravity conjecture,''
  JHEP {\bf 0603}, 094 (2006)
  [arXiv:hep-th/0601137].

\bibitem{Adams:2006sv}
  A.~Adams, N.~Arkani-Hamed, S.~Dubovsky, A.~Nicolis and R.~Rattazzi,
  ``Causality, analyticity and an IR obstruction to UV completion,''
  JHEP {\bf 0610}, 014 (2006)
  [arXiv:hep-th/0602178].

\bibitem{Ooguri:2006in}
  H.~Ooguri and C.~Vafa,
  ``On the geometry of the string landscape and the swampland,''
  arXiv:hep-th/0605264.

\bibitem{Li:2006jj}
  M.~Li, W.~Song, Y.~Song and T.~Wang,
  ``A weak gravity conjecture for scalar field theories,''
  arXiv:hep-th/0606011.

\bibitem{Kats:2006xp}
  Y.~Kats, L.~Motl and M.~Padi,
  ``Higher-order corrections to mass-charge relation of extremal black holes,''
  arXiv:hep-th/0606100.

\bibitem{Medved:2006ht}
  A.~J.~M.~Medved,
  ``An implication of 'gravity as the weakest force',''
  arXiv:hep-th/0611196.

\bibitem{Gasperini:2006as}
  M.~Gasperini,
  ``A new scale in the sky,''
  arXiv:hep-th/0611227.

\bibitem{Guth:1980zm}
  A.~H.~Guth,
  ``The Inflationary Universe: A Possible Solution To The Horizon And Flatness
  Problems,''
  Phys.\ Rev.\ D {\bf 23}, 347 (1981).

\bibitem{Guth:1982pn}
  A.~H.~Guth and E.~J.~Weinberg,
  ``Could The Universe Have Recovered From A Slow First Order Phase
  Transition?,''
  Nucl.\ Phys.\ B {\bf 212}, 321 (1983).

\bibitem{Albrecht:1982wi}
  A.~Albrecht and P.~J.~Steinhardt,
  ``Cosmology For Grand Unified Theories With Radiatively Induced Symmetry
  Breaking,''
  Phys.\ Rev.\ Lett.\  {\bf 48}, 1220 (1982).

\bibitem{Linde:1982zj}
  A.~D.~Linde,
  ``Coleman-Weinberg Theory And A New Inflationary Universe Scenario,''
  Phys.\ Lett.\ B {\bf 114}, 431 (1982).

\bibitem{Robinson:2005fj}
  S.~P.~Robinson and F.~Wilczek,
  ``Gravitational correction to running of gauge couplings,''
  Phys.\ Rev.\ Lett.\  {\bf 96}, 231601 (2006)
  [arXiv:hep-th/0509050].

\bibitem{Linde:1983gd}
  A.~D.~Linde,
  ``Chaotic Inflation,''
  Phys.\ Lett.\ B {\bf 129} (1983) 177.

\bibitem{Linde:1993cn}
  A.~D.~Linde,
  ``Hybrid inflation,''
  Phys.\ Rev.\  D {\bf 49}, 748 (1994)
  [arXiv:astro-ph/9307002].

\bibitem{Liddle:1998jc}
  A.~R.~Liddle, A.~Mazumdar and F.~E.~Schunck,
  ``Assisted inflation,''
  Phys.\ Rev.\ D {\bf 58}, 061301 (1998)
  [arXiv:astro-ph/9804177].

\bibitem{Dimopoulos:2005ac}
  S.~Dimopoulos, S.~Kachru, J.~McGreevy and J.~G.~Wacker,
  ``N-flation,''
  arXiv:hep-th/0507205.

\bibitem{Zlatev:1998tr}
  I.~Zlatev, L.~M.~Wang and P.~J.~Steinhardt,
  ``Quintessence, Cosmic Coincidence, and the Cosmological Constant,''
  Phys.\ Rev.\ Lett.\  {\bf 82}, 896 (1999)
  [arXiv:astro-ph/9807002].


\end{thebibliography}
\end{document}